\documentclass[iop,twocolumn,showpacs,preprintnumbers,amsmath,amssymb,superscriptaddress]{revtex4}

\usepackage{graphicx}
\usepackage{dcolumn}
\usepackage{bm}

\begin{document}

\title{Efficiency of informational transfer in regular and complex networks}

\author{I. Vragovi\'c}

\affiliation{Departamento de F\'isica Aplicada,
Instituto Universitario de Materiales and Unidad Asociada CSIC-UA,
Universidad de Alicante, E-03080 Alicante, Spain}

\author{E. Louis}

\affiliation{Departamento de F\'isica Aplicada,
Instituto Universitario de Materiales and Unidad Asociada CSIC-UA,
Universidad de Alicante, E-03080 Alicante, Spain}

\author{A. D\'iaz-Guilera}

\affiliation{Departament de F\'isica Fonamental,
Universitat de Barcelona, E-08028 Barcelona, Spain}

\date{\today}

\begin{abstract}
We analyze the process of informational exchange through complex networks by measuring network efficiencies. Aiming to study non-clustered systems, we propose a modification of this measure on the local level. We apply this method to an extension of the class of small-worlds that includes {\it declustered} networks, and show that they are locally quite efficient, although
their clustering coefficient is practically zero. Unweighted systems with small-world and scale-free topologies are shown to be both globally and locally efficient. Our method is also applied to characterize weighted networks. In particular we examine the properties of underground transportation systems
of Madrid and Barcelona and reinterpret the results obtained for the Boston subway network.

\end{abstract}

\maketitle

\section{Introduction}
\vspace{2mm}

Modelling of complex systems as networks of coupled elements, such as chemical systems \cite{scala2000,amaral2000}, neural networks \cite{lago2000}, epidemiological \cite{pandit1999,kuperman2001a} and social networks \cite{liljeros2001} or the Internet \cite{faloutsos1999},
has been a subject of intense study in the last decade.
Networks can be classified into three broad groups: i) regular networks, ii) random networks, and iii) systems of complex topology, including small-world \cite{wattsstrogatz1998,watts1999}
and scale-free networks
\cite{liljeros2001,albert1999,barabasi1999,newman2001,albertbarabasi2001}. In addition, networks can be  unweighted or weighted, depending on whether links are equal or different. Weights can be: physical distances, times of propagation of informational packets, inverse velocity of chemical reactions,
strength of interactions, etc. \cite{gopal2001,kuperman2001,moukarzel2002,graham2003}.

Commonly used regular networks are square or cubic lattices, both having squares as basic cycles \cite{graham2003,cartwright2000,degli2002}. Aiming to describe clustering in social networks, i.e. to account for triangles of connected nodes as basic cycles \cite{wassermanfaust94}, clustered rings  were introduced, in which each site was linked to all its neighbors from the first up to the $K$-th \cite{pandit1999,gopal2001,gao2001,hong2002}. The study of random graphs \cite{erdosrenyi59,erdosrenyi60,erdosrenyi61,bollobas85} was motivated by the observation of real networks that often appeared to be random. Complex networks having a topology in between those of random and regular networks were later introduced.  An outstanding example is the  small-world model \cite{wattsstrogatz1998,milgram1967}. Small worlds are constructed by randomly rewiring links of a regular graph \cite{wattsstrogatz1998} (so that the number of links remains constant, while the structure is changed) or adding new links to it \cite{newmanwatts1999b} (changing both the structure and the
number of links) with a probability $p$. In this way, shortcuts between distant nodes are created. The rewiring/adding probability $p$ indicates, on average, the degree of disorder of the network (it varies from $p=0$ for a regular up to $p=1$ for a random graph). Small-worlds are highly clustered showing triangles of nodes like regular networks, while having  small distances between sites as in random systems \cite{pandit1999,kuperman2001a,zanette2002,bagnoli2001,moukarzel1999}. Recently, it was realized that many social and biological networks had a degree (connectivity) distribution that was  not
Poisson-like, as in random and small-world networks, but rather
a power law. Such systems were called scale-free \cite{albert1999,barabasi1999,newman2001,albertbarabasi2001} and are continuously growing open systems constructed by attaching  new nodes preferentially  to  nodes of higher degree \cite{barabasi1999}. Various modifications of this basic procedure have been proposed: nonlinear preferential attachment \cite{krapivsky2000}, initial attractiveness \cite{dorogovtsev2000}, and aging of sites and degree constraints
\cite{albertbarabasi2001} or node fitness \cite{bianconi2001}. Moreover, introducing a finite memory of the nodes, large highly clustered systems can be obtained, representing a combination of scale-free networks and regular lattices \cite{klemm2002}.

\begin{figure}[h]
\includegraphics[angle=0,width=9cm]{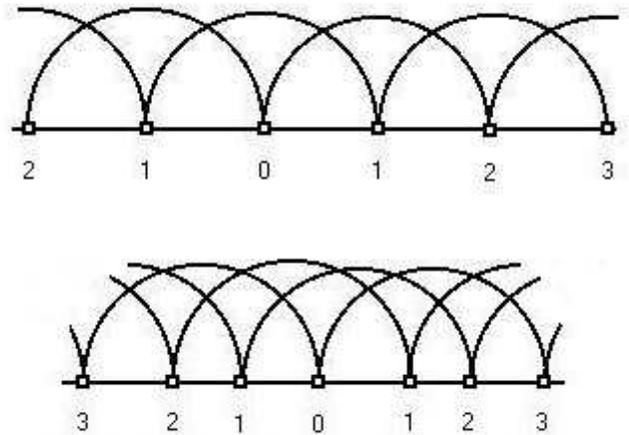}
\caption{Illustrates the link structure in clustered rings (upper) with
connections
to the second  nearest-neighbors ($K=2$), and declustered rings (lower) with connections to the third nearest-neighbors ($K=\bar{3}$).}
\label{cldc}
\end{figure}

Our aim is to compare the efficiency of informational transfer on the regular and complex networks described above. In Section \ref{netchar} we describe the  networks and define the quantities used to characterize them. In \ref{nettypes} an extension of the class of small-world, referred to as {\it declustered}, is proposed. In \ref{seceff} we discuss the efficiency measures reported in the literature and propose the alternatives required to handle non-clustered systems. Some of the new measures defined here are an extension of those reported in \cite{latora2001}. Section \ref{char} is devoted to discuss the properties of various unweighted networks.  Introducing physical distances, efficiencies  of weighted networks are defined in Section \ref{weight} and used to examine underground transportation systems. Our achievements  are summarized in Section \ref{concl}.

\section{Methods}
\label{netchar}

\subsection{Types of Networks}
\label{nettypes}

The networks  analyzed here are:  clusters of the square lattice, clustered and declustered  regular rings, as examples of regular systems; clustered Watts-Strogatz and declustered small-worlds, and ordinary Albert-Barab\'asi scale-free networks, representing complex systems. All networks are chosen so that the ratio between the number of links $N_l$ and the number of sites
$N$ is kept constant $N_l/N=2$ (this gives an average connectivity $<k>=4$).

Concerning regular two-dimensional networks, calculations were performed for $l \times l$ clusters of the square lattice, with periodic boundary  conditions: $\mbox{node}(i+l,j)=\mbox{node}(i,j)$ and $\mbox{node}(i,j+l)=
\mbox{node}(i,j)$.  In the case of regular rings, we analyze the simplest clustered lattices with additional connections only to the next-nearest neighbors ($K=2$) \cite{zanette2002,huang2003}, see Fig. \ref{cldc}. In addition, we study rings with a zero clustering coefficient constructed  by adding links from each site to {\it only} its $n$-th neighbors. We call them
{\it
declustered regular ring networks} and designate their coordination parameter as $K=\bar{n}$, (shown also in Fig. \ref{cldc}). Therefore, in our notation $K=n$ means that each site is additionally linked to {\it all} of its ring neighbors
from the second to the $n$-th, while $K=\bar{n}$ implies that only links to its $n$-th neighbors are added. For such declustered networks, basic loops are squares for any $\bar{n}$, with edges on sites $i$, $i+n$, $i+1$ and $i+n+1$.
Our motivation to analyze networks with a negligible clustering comes from the fact that such systems can be quite often found in nature or artifacts (for instance in transportation underground networks). Such networks are usually very sparse with $N_l \approx N$ \cite{latora2001}.

We  differentiate between ordinary clustered small-world and {\it
declustered} small-world, depending on the initial regular network. We will construct small-world networks starting from clustered and declustered regular  networks with $K=2$ and $K=\bar{3}$, respectively. Moreover, as our focus is on the
effects of network topology, we compare networks with the same links to size ratio. Thus, shortcuts are created by randomly {\it rewiring} links between each site and its more distant neighbors with probability $2p \leq 1$, while connections to the nearest neighbors are kept unchanged. In this way, the ring structure is preserved and the problem of disconnected graphs is avoided
\cite{newmanwatts1999b}. The total number of rewired links would approach $pN \leq N/2$ for large $N$. Finally, we construct scale-free networks starting with a fully connected graph
of $m_0=5$ nodes and $n_0=10$ links. At each step a new node is added, with $m=2$ edges to the old nodes, so that the ratio $N_l/N=2$ is kept constant.

\begin{figure}[h]
\includegraphics[angle=270,width=9cm]{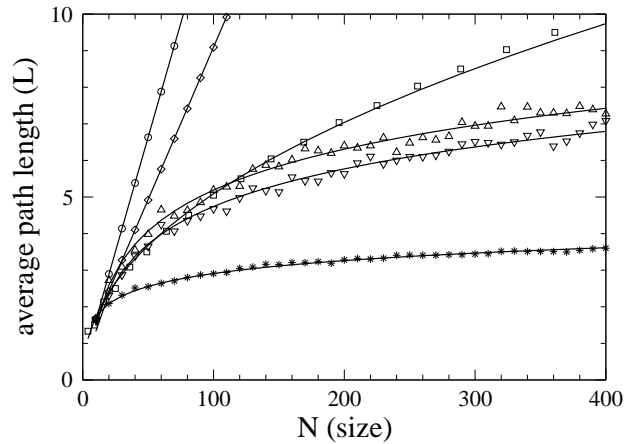}
\caption{Average path length versus network size ($N$) for the networks investigated in this work (all with average connectivity $\langle k \rangle=4$). Regular rings: clustered $K=2$ (circle), declustered $K=\bar{3}$ (diamond) and clusters of the square lattice (square). Complex: small-worlds with a probability of rewiring of $p = 10 \%$ - clustered $K=2$ (triangle up), declustered $K=\bar{3}$ (triangle down); scale-free networks with $m_0=5$ and $m=2$ (star). Lines are fits of the numerical results in the range $N=10-400$: clustered ring $L=0.13N+0.39$, declustered ring $L=0.083N+0.77$, cluster of the square lattice $L=0.59N^{0.47}$, clustered small world $L=1.61 {\rm ln} N - 2.24$, declustered small world $ L=1.48 {\rm
ln} N - 2.08$ and scale-free network $L = 2.16 {\rm ln ln} N - 0.32$.}
\label{lengthregcompl}
\end{figure}

\subsection{Average path length and clustering coefficient}
\label{pathclust}

The structural properties of a graph are usually quantified by the average path length $L$ and the clustering coefficient $C$
\cite{wattsstrogatz1998,albertbarabasi2001}. The average path length is calculated as the network average of the shortest graph distances between two nodes ($d_{ij}$) for all possible pairs:

\begin{equation}
L = \frac{1}{N(N-1)} \sum_{i \neq j} d_{ij},
\end{equation}
defined for connected graphs for which all $d_{ij}$ are finite.

The clustering coefficient $C$ measures to which extent are neighbors of each site connected to each other. It is calculated as a network average:

\begin{equation}
C = \frac{1}{N} \sum_{i} C_{i}.
\end{equation}
where $C_{i}$ is the ratio of the existing number of links between the neighbors of a site $i$ and the maximum possible number of them $k_i[k_i-1]/2$; $k_i$ being its connectivity (degree).

It is worth noting that, although the clustering coefficient of almost all real networks is very high \cite{albertbarabasi2001}, it seems that it is much less important for the collective dynamical behavior of a network than the average path length. Moreover, in some cases, like square lattices or declustered
ring networks, the usual clustering coefficient fails to correctly quantify the underlying order of the hierarchical structure of the system. Recently it  was proposed that such grid-like structures should be characterized by a grid-coefficient, numbering the fraction of all the loops of length 4 ({\it quadrilaterals}) passing through each node \cite{cm0212026,albertnewpaper}. Analysis of real networks, such as Internet, Web and scientific coautorship, reveals a good local rectangular clustering \cite{cm0212026}. However, similarly to ordinary clustering coefficient based exclusively on triangles,
this new measure concentrates only on square loops. Any attempt to analyze more sparse networks with longer basic cycles would call for the introduction of new coefficients of even higher orders. It would be much more useful to find a single measure of local properties that could be applied to any type of
networks. Furthermore, it is not clear what is the physical meaning of these various coefficients and how would they be related to the dynamical behavior of the network.

\subsection{Efficiency of informational exchange}
\label{seceff}
Another approach to analyze global and local properties of a network is introducing the concept of efficiency of informational exchange through the network \cite{latora2001,crucitti2004,li2004}.

\subsubsection{Global efficiency}
\label{gleff}

We assume that it is easier to transfer information from one site to another if they are closer to each other. Therefore, the efficiency in the communication between two sites $i$ and $j$ is calculated as the inverse of the shortest path length $d_{ij}$ between these two sites: $\epsilon_{ij}=1/d_{ij}$. Contrary
to the average path length, efficiency can be determined even if there is no path between $i$ and $j$, as in the case of disconnected graphs: $\lim_{d_{ij} \rightarrow \infty} \epsilon_{ij} = 0$. The global efficiency of the
network is calculated as the average over all pair of nodes
\cite{latora2001}:

\begin{equation}
E^{g}= \frac{1}{N(N-1)} \sum_{i \neq j} \frac{1}{d_{ij}},
\label{effglob}
\end{equation}
and is normalized to its possible largest value $N(N-1)$, for totally connected graph having $N(N-1)/2$ edges. Physically, $E^{g}$ measures the efficiency of a system with parallel exchange of information, while $1/L$ accounts for the
efficiency of a sequential propagation of a single informational packet along the network. In the case of real networks, $E^{g}$ gives a better measure for the transfer of information than $1/L$, although quite often $1/L$ could be a reasonable approximation of $E^{g}$ \cite{latora2001}.

\begin{figure}[h]
\includegraphics[angle=270,width=9cm]{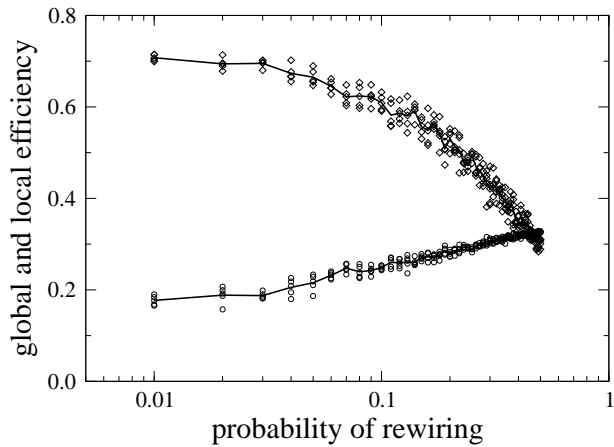}
\caption{Global (circle) and local (diamond) efficiencies versus the rewiring parameter $p$ for clustered small-worlds. The initial regular ring was  clustered with $K=2$.}
\label{EffGLcl}
\end{figure}

\subsubsection{Local efficiency}
\label{loceff}

A similar  definition can be implemented on a local level. As a counterpart of the  clustering coefficient $C$, the local efficiency could be defined as an average efficiency of the local subgraphs of the first neighbors ($j,k \in \Gamma_{i}$) of each site $i$ \cite{latora2001}:

\begin{equation}
E^{l}_{0}= \frac{1}{N} \sum_{i}
\frac{1}{k_i[k_i-1]} \sum_{j \neq k \in \Gamma_1} \frac{1}{d^0_{jk/i}}.
\end{equation}
Here $d^0_{jk/i}$ is the shortest path length between sites $j$ and $k$ passing only through other elements of that local subgraph of neighbors ($\Gamma_1$), which is indicated by superscript 0. In such a way, the clustering coefficient is equal to the local efficiency when only direct connections between $j$ and $k$ are considered.

We propose a new definition of local efficiency, taking into account that neighbors of each reference site $i$ can actually exchange information along paths including sites which do not necessarily belong to the local subgraph of $i$'s neighbors ($m \notin \Gamma_1$). In order to measure the efficiency of
communication between the nearest-neighbors of $i$ when it is removed, we must only exclude site $i$ from such a path ($d_{jk/i}$):

\begin{equation}
E^{l}_1 = \frac{1}{N} \sum_{i}
\frac{1}{k_i[k_i-1]} \sum_{j \neq k \in \Gamma_1} \frac{1}{d_{jk/i}}.
\end{equation}
When applying such a concept on graphs without triangle cycles, we will see that they can transfer the information quite efficiently on a local level, although their clustering coefficient is zero. It is worth noting that in the
definition of \cite{latora2001} (see Eq. (5)) local efficiency depends only on the links present in the graph $\Gamma_1$ of the first neighbors of site $i$. It is calculated excluding both site $i$ and the rest of the network ($m \notin \Gamma_1$). In the new definition, however, local efficiency depends on the full network topology and is calculated cutting off only site $i$.

\begin{figure}[h]
\includegraphics[angle=270,width=9cm]{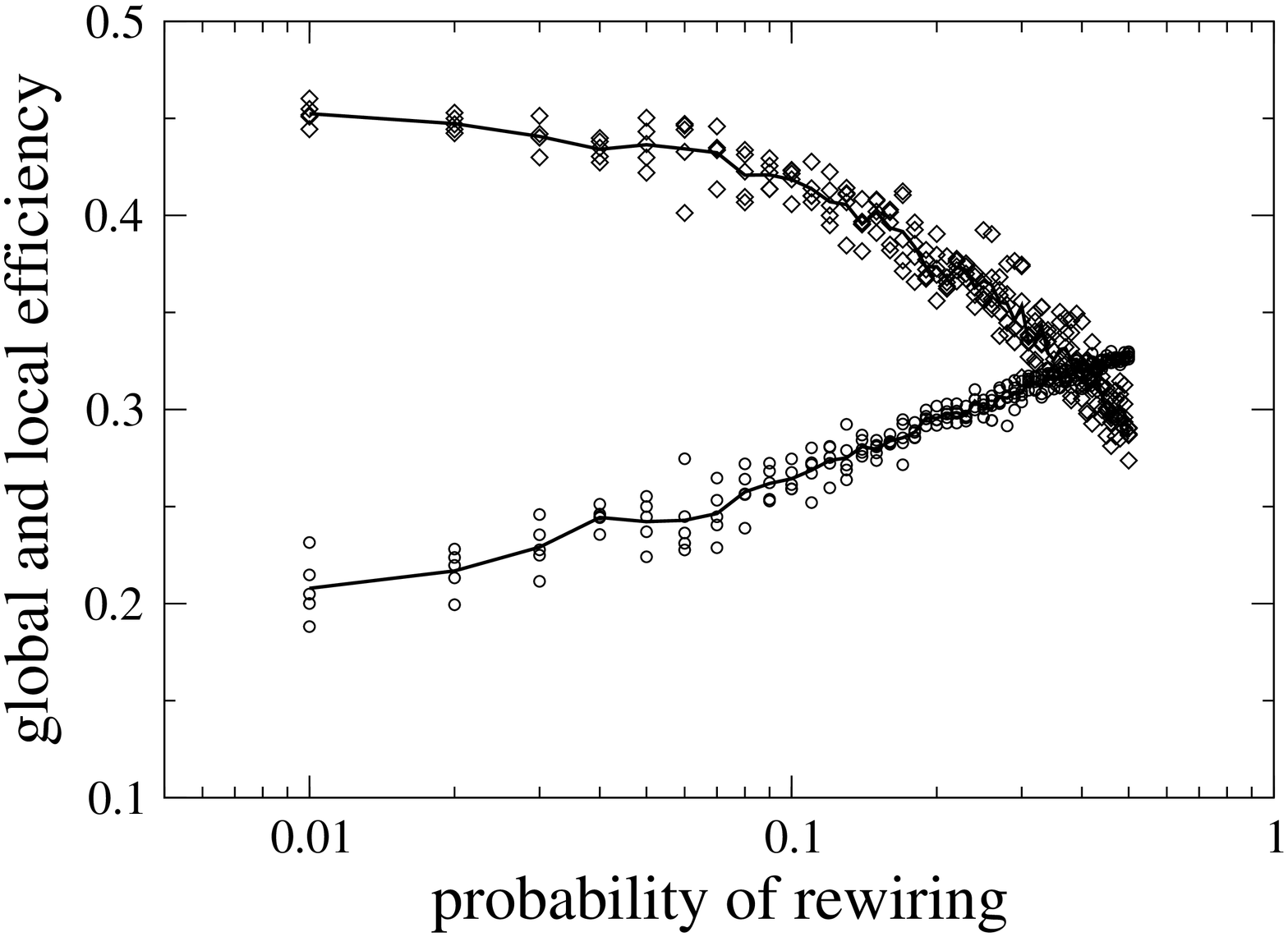}
\caption{Global (circle) and local (diamond) efficiencies versus the rewiring parameter $p$ for declustered small-worlds. The initial regular ring was declustered with $K=\bar{3}$.}
\label{EffGLdc}
\end{figure}

The clustering coefficient was introduced to measure the closeness of sites or the locality of a network \cite{wattsstrogatz1998}. Locality tells us up to what
extent the neighbors of a site remain close to each other after this site is cut off. Regular networks are precisely those which show the highest locality.  The criteria used for the calculation of the clustering coefficient takes direct connections as a substitute for closeness. From our standpoint, it is not
necessary to have two sites directly connected in order to conclude that they are close to each other. They will be far away only if the length of the path that connects them (going through the rest of the network, except site $i$) turns out to be large. In this way, the level of closeness (or locality) among neighbors of $i$ depends on network topology. At first sight, our measure of
locality mixes global and local properties. However, we must note that the path between sites that are close to each other does not go throughout the whole remaining network, but only through the close surroundings of these sites. The immediate surroundings of these sites will overlap in a great extent,  defining a common local region. Therefore, our definition of local efficiency is
logically consistent, as it depends mainly on local topology.

\section{Characterization of unweighted networks}
\label{char}

\subsection{Average path length and clustering coefficient}

Fig. \ref{lengthregcompl} shows the  average path length $L$ for different types of networks with a link-to-size ratio $N_l/N=2$, versus the size of  the system $N$. Fittings of the numerical results are given in the caption of  the figure. In ordered rings, $L$ scales linearly with size ($L_{\rm ring} \sim
N $), with a slope that is smaller for the declustered network $K=\bar{3}$ (the slopes given in the figure caption are close to the exact results, namely, 1/8 and 1/12, respectively, see \cite{lengths}). For square lattices, the dependence is
sublinear, i.e. $L_{\rm square} \approx \sqrt{N}/2$ \cite{lengths}. Random graphs, in its turn, are known to obey a logarithmic scaling ($L_{\rm rand}  \sim \ln N$) \cite{albertbarabasi2001}. Such a behavior is also observed in the case of Watts-Strogatz small-world. As Fig. \ref{lengthregcompl} clearly shows declustered small-world behaves qualitatively in the same way as Watts-Strogatz networks, in both cases the average path length is proportional
to ${\rm ln}N$. The average path length in declustered small-world is shorter than in the standard small-world, as edges of basic square cycles of the initial declustered network couple more distant sites. Finally, scale-free systems appear to be
ultra-small \cite{cohen2003}, with a double logarithmic scaling  $L_{sf} \propto \ln \ln N$ (see Fig. \ref{lengthregcompl}).

In order to differentiate between random graphs and small-worlds, both  having the same scaling of the average path length and a Poisson distribution of degrees, the clustering coefficient is used. Switching from highly clustered regular graphs to small-worlds by introducing a few shortcuts does not significantly alter the clustering coefficient
\cite{wattsstrogatz1998,albertbarabasi2001}. It remains quite large up to high values of the rewiring parameter $p$. For $p \approx 1$ most triangle loops  are broken, leading to random graphs with negligible values of $C$. The small-world behavior shows up at small $p$, when both the average path length and the clustering coefficient have large values
\cite{wattsstrogatz1998,albertbarabasi2001}.

\subsection{Efficiencies}

The aforementioned criteria identify declustered small-worlds as random
networks. But, is this actually the case?. Applying the concept of global
and
the redefined local efficiency, we clearly identify the crucial differences
between various networks. The clustered regular ring-lattice with $K=2$ is
locally very efficient $E^{l}_1=0.722$, due to its high clusterization, see
Table
\ref{table1}. Global efficiency is quite low $E^{g}(N=100)=0.154$, which
corresponds to a long average path length $L(100)=12.88$. From our
standpoint, a declustered ring-lattice with $K=\bar{3}$ has the same
characteristics. Local efficiency $E^{l}_1=0.458$ is relatively good, 
although
the clustering coefficient and the originally proposed local measure of
efficiency \cite{latora2001} are both zero. Globally, we obtain slightly a
larger
value of $E^{g}(100)=0.188$ (or shorter $L(100) = 9.09$), due to the 
presence
of longer range links. Therefore, regular rings are in general locally 
efficient
and globally inefficient.

\begin{table}[h]
\caption{Average path length, clustering coefficient, and global and local
efficiencies for homogeneous networks.}
\begin{tabular}{|l|rrrr|} \hline
& $L$ & $C$ &  $E^{g}$ & $E^{l}_1$  \\ \hline
regular clustered & \hspace{1mm} 12.88 & \hspace{1mm} 0.5 & \hspace{1mm}
0.154 & \hspace{1mm} 0.722 \\
regular declustered & 9.09 & 0 & 0.188 & 0.458 \\
2D square & 5.05 & 0 & 0.258 & 0.417 \\
random & 3.40 & 0.02 & 0.328 & 0.280 \\ \hline \hline
\end{tabular}
\label{table1}
\end{table}

\begin{figure}[h]
\includegraphics[angle=270,width=9cm]{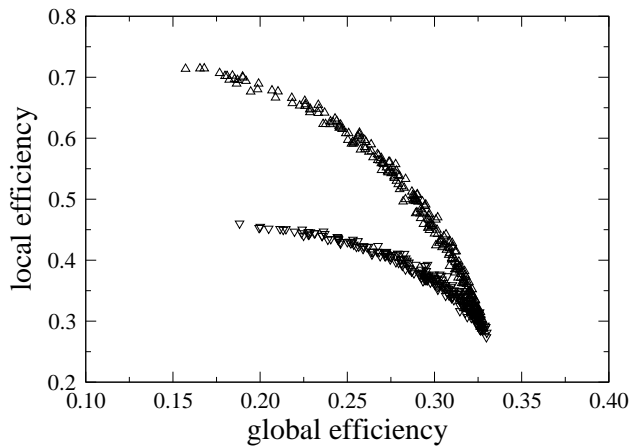}
\caption{Global versus local efficiencies for clustered (triangle up) and
declustered (triangle down) small-worlds. The initial regular rings were
clustered ($K=2$, $E^g=0.258$ and $E^l=0.722$) and declustered
($K=\bar{3}$, $E^g=0.188$ and $E^l=0.458$). A rewiring parameter of 0.5
leads to a single random graph.}
\label{EffGL}
\end{figure}

Introducing a small number of shortcuts into a regular graph to produce a
small-world network, does not significantly alter its local topology and 
local
efficiency. On the other hand, global efficiency is appreciably improved. As
illustrated in Fig. \ref{EffGLcl} and \ref{EffGLdc} this is valid for both
clustered
and declustered small-worlds. The rewiring parameter $p$ was varied in the
range 0.01-0.5 (for each value of $p$ results for five graph realizations 
are
shown). In a random graph (large $p$), the efficiency on the global scale
becomes even better, but local efficiency is  strongly deteriorated. The
distinction between regular (left end), small-world (middle part) and random
networks (right end of curves) is clearly depicted in Fig. \ref{EffGL}. We 
can
conclude that small-worlds are both {\it globally} and {\it locally 
efficient}
\cite{latora2001}.

\begin{figure}[h]
\includegraphics[angle=270,width=9cm]{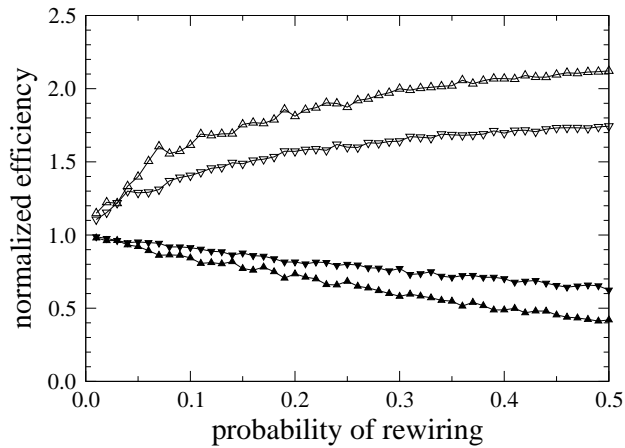}
\caption{Global (empty symbols) and local (filled symbols) efficiencies
normalized to the value of the initial regular ring versus the rewiring
parameter
$p$, for clustered (triangle up) and declustered (triangle down) 
small-worlds.}
\label{EffGLnorm}
\end{figure}

Normalizing the global efficiency of a given small-world to the values of 
the
initial ring (see Fig \ref{EffGLnorm}), we note that it is improved in a
relatively
better way in the case of clustered networks. This is due to the fact that
initially
there are only short $K=2$ links to be rewired into links of longer range.
Starting from a declustered regular ring $K=\bar{3}$, links can eventually 
be
rewired into shorter $K=2$ links, that do not increase the global 
efficiency.
The
same type of normalization can be done for the local efficiencies as shown 
in
cp. Fig \ref{EffGLnorm}. Now, possible rewiring of $K=\bar{3}$ links into 
$K=2$
links actually improves local efficiency, assuring that normalized values 
for
declustered small-worlds are always larger than values for the clustered
network.

\subsection{The Size of the Network}

\begin{figure}[h]
\includegraphics[angle=270,width=9cm]{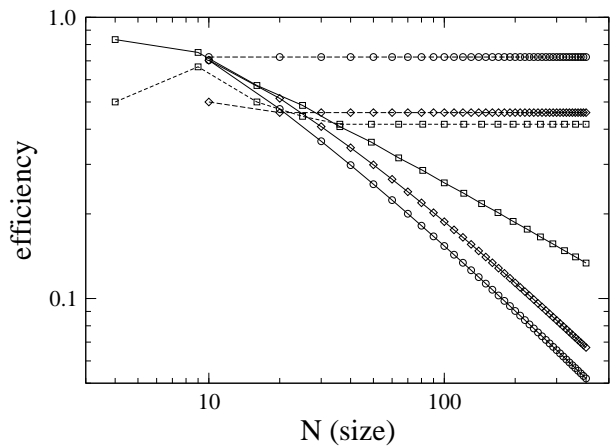}
\caption{Global (full line) and local (dashed line) efficiencies versus 
network
size ($N$) for  regular networks with a constant connectivity ($k_i=4$):
clustered $K=2$ (circle), declustered $K=\bar{3}$ (diamond) and clusters of
the square lattice (square).}
\label{sizereglog}
\end{figure}

\begin{figure}[h]
\includegraphics[angle=270,width=9cm]{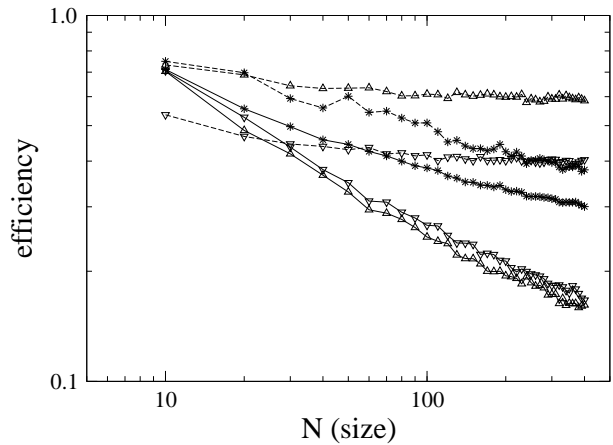}
\caption{Global (full line) and local (dashed line) efficiencies versus the
network
size ($N$) of complex networks with a constant average connectivity 
($\langle
k \rangle=4$). Small-worlds with a  rewiring parameter $p = 0.1$: clustered
$K=2$ (triangle up), declustered $K=\bar{3}$ (triangle down). Scale-free
networks with $m_0=5$ and $m=2$ (star).}
\label{sizecompllog}
\end{figure}

The dependence of global and local efficiencies on network size (see Fig.
\ref{sizereglog} and Fig. \ref{sizecompllog}) shows that regular networks 
have
a local efficiency  that does vary with $N$ (i.e. unchanged local topology). 
The
global efficiency decreases with size, with the minimal
$\epsilon^{min}_{ij}=1/d^{max}_{ij}$ scaling as $1/N$. In the case of
small-worlds, the local topology is not much affected by the presence of a 
few
shortcuts, leading again to a approximately constant local efficiency. 
Global
efficiency is now expected to decrease at a lower rate, because the minimal
$\epsilon^{min}_{ij}=1/d^{max}_{ij}$ scales as $1/\ln N$. The results are 
quite
different for scale-free networks. While efficiency on a global level 
decreases
with slower rate with respect to the other networks, local efficiency is
significantly decreased. The reason is that the clustering coefficient, 
giving
the
main contribution to the local efficiency, decreases exponentially with the 
size
of the ordinary scale-free system \cite{albertbarabasi2001}. We expect that 
the
local efficiency of highly clustered scale-free networks \cite{klemm2002} 
would
be mainly independent of the network size, as their clustering coefficient
approaches a high stationary value already for $N \sim 10^2$. Such a
tendency is observed in scale-free Internet networks \cite{pastor2001}, 
where
the clustering coefficient even increases over consecutive years. From our
standpoint,  Internet grows keeping constant local efficiency.

\begin{figure}[h]
\includegraphics[angle=270,width=9cm]{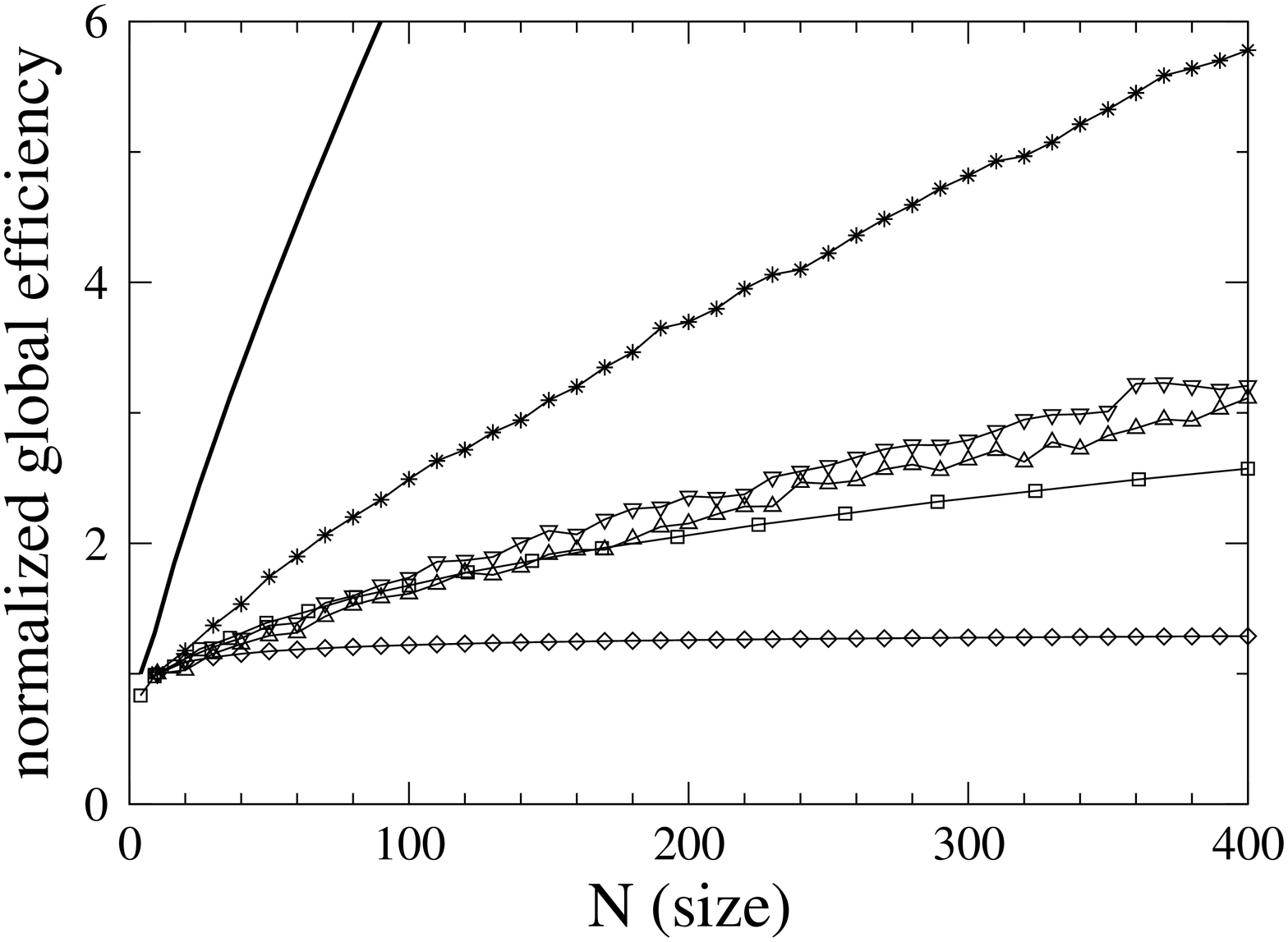}
\caption{Normalized global efficiency versus network  size ($N$) with a
constant average connectivity ($\langle k \rangle=4$). The global efficiency 
of
each particular network is normalized to the value for the regular clustered
$K=2$ network of the same size. Regular: declustered $K=\bar{3}$ (diamond)
and clusters of the square lattice (square). Complex: small-worlds with a
rewiring parameter $p = 0.1$ - clustered $K=2$ (triangle up), declustered
$K=\bar{3}$ (triangle down); scale-free networks with $m_0=5$ and $m=2$
(star). Fully connected graph: full line.}
\label{effglobregcomplnorm}
\end{figure}

The reason for the  observed decrease of global efficiency is related to the
fact
that, for a constant ratio $N_l/N=a$, increasing the size produces gradually
more sparse graphs with longer average path length. The total number of
possible links is given by $N(N-1)/2$, while the number of links actually
present
in the system is $N_l=a N$. This leads to a decrease of the density of links 
as
$\eta=2a/(N-1)$.

\subsection{Normalized global efficiency and basic network}

We can normalize the results for each type of network to the values for
clustered regular $K=2$ rings of the same size $e^{g}=E^{g}/E^{g}_{r}$. The
results are reported in Fig. \ref{effglobregcomplnorm}. The normalized 
global
efficiency of a declustered regular network is slightly larger, but does not
change with size. On the contrary, it increases with size for  the square
lattice,
small-world and scale-free networks. This normalization is necessary if we
want to examine how a pure change of topology improves transfer of
information, without addition of new links. Eq. (\ref{effglob}) tells us how
efficient is a network on a global scale relatively to the ideal case of a 
fully
connected graph. Such a comparison can be misleading, because does not
take into account that graphs are commonly sparse. Increasing the size, 
while
keeping the $N_{l}/N$ ratio constant, global efficiency decreases for any 
kind
of sparse networks. In contrast to a fully connected graph, each type of
network would be seen as inefficient, no matter what is the underlying
topology. Therefore our opinion is that a particular network should be
compared with a corresponding basic network with the same number of sites
$N$ and links $N_l$. The basic network is a periodic system with the longest
possible average path length or the smallest possible global efficiency for
given
$(N,N_l)$. It can be constructed in the following way:

a) Start from an initial standard ring of $N$ sites and  $N$ links.

b) Add links between each site and its closest surrounding sites (the
next-nearest neighbors, the next-next-nearest neighbors, and so on), up to 
all
$N_l$ links are used. The result is a $K=N_{l}/N$ regular ring.

c) In case that the ratio $N_l/N$ is not an integer, the last set of $n_l < 
N$
links should be evenly distributed among the sites.

\noindent
In such a way, complex systems such as small-worlds or scale-free networks,
are identified to be globally efficient in comparison with the corresponding
inefficient basic (regular) networks.

\section{Weighted networks}
\label{weight}

\subsection{Efficiency measures}

In this section we focus on a particular type of weighted graphs, where
physical distances are introduced. A real network is described by both the
connectivity matrix and the matrix of physical distances \cite{latora2001}. 
The
shortest physical path length $\tilde{d}_{ij}$ between two sites $i$ and $j$ 
is
the path with the smallest sum of distances, no matter the number of links 
the
network has. Only in the case of links of equal lengths ($\lambda$), the
physical and the graph shortest paths coincide, i.e. $\tilde{d}_{ij}= 
\lambda
d_{ij}$. The efficiencies of a real network could be calculated using the
formulas given in Sec. \ref{seceff}, replacing $d_{ij}$ by $\tilde{d}_{ij}$. 
In
order
to keep these quantities dimensionless, a suitable normalization should be
performed. The originally proposed efficiency measures \cite{latora2001} are
normalized to the values for the fully connected graph of the same size:

\begin{equation}
\tilde{E}^{g}_1= \displaystyle \frac{\displaystyle \sum_{i \neq j}
\frac{1}{\tilde{d}_{ij}}}{\displaystyle \sum_{i \neq j} \frac{1}{l_{ij}}},
\label{effglobtilde1}
\end{equation}
where $l_{ij}$ is a physical distance (or length of a possible direct link)
between
sites $i$ and $j$. We propose a slightly different measure, that gives 
similar
quantitative results. Instead of comparing the network as a whole with the 
ideal
graph, we analyze the efficiency of each particular shortest path
$\tilde{d}_{ij}$
separately:

\begin{equation}
\tilde{E}^{g}_2= \frac{1}{N(N-1)} \sum_{i \neq j} 
\frac{l_{ij}}{\tilde{d}_{ij}}.
\label{effglobtilde2}
\end{equation}
The main reason for using this network average of path efficiencies is that 
in
weighted networks the shortest paths going through a lot of sites can very
often be as efficient as direct links between pairs of sites, significantly
contributing to the network efficiency. If we neglect possible delays 
between
received and subsequent emitted  information, we see that the straight  path
going through many sites is the same as a direct straight link between two 
end
nodes. The weighted efficiency of such a straight-path regular graph will be 
the
same as that of the corresponding fully connected system. This simple
example raises a question: is it necessary at all to impose a small-world
topology in order to achieve a higher global efficiency in a weighted 
network?.
A closer look  into a $K=1$ weighted ring shows that the weighted efficiency 
of
the longest path between two opposite sites is very high, i.e.
$l_{ij}/\tilde{d}_{ij}
\approx 2R/(R \pi) = 2/\pi$. That is, 64 \% of the efficiency of the direct
link!.
For sites closer to each other or for a regular network with $K>1$, this 
ratio
is
even larger. This result is not surprising, as we assume that the speed of
informational transfer through all the links is constant. Therefore links
between
faraway sites do not represent shortcuts, because the time needed to 
transfer
the information increases with the physical length of the link. The shortcut
would be created if the transfer is instantaneous or at least very fast, so 
that
the corresponding transfer time is much shorter than the characteristic 
times of
the underlying dynamics. Such a case would be a flight of an infected person
by an airplane, when the basic mechanism is a slow spreading of the disease
through direct contacts \cite{pandit1999}, but definitely not the 
transportation
systems (like railways) where the speed of all the vehicles is limited and
usually constant.

Similar measures can  be defined on the local scale. The original weighted
local efficiency, when paths are going only through the first 
nearest-neighbors
of a reference site, is given by \cite{latora2001}:

\begin{equation}
\tilde{E}^{l}_{0}= \displaystyle \frac{\displaystyle  \sum_{i} \displaystyle
\frac{1}{k_i[k_i-1]} \displaystyle  \sum_{j \neq k \in \Gamma_1} 
\displaystyle
\frac{1}{\tilde{d}^0_{jk/i}} }{ \displaystyle
\sum_{i} \displaystyle  \frac{1}{k_i[k_i-1]} \displaystyle  \sum_{j \neq k 
\in
\Gamma_1} \displaystyle  \frac{1}{l_{jk}} }.
\end{equation}
Allowing for paths going through the rest of the network we define
$\tilde{E}^{l}_{1}$ by simply replacing $\tilde{d}^0_{jk/i}$ by
$\tilde{d}_{jk/i}$.
Finally, we can again normalize each path separately, instead of normalizing
the whole sum by the value for fully connected network 
($\tilde{E}^{l}_{2}$).

\subsection{Analysis of subway transportation systems}

Underground transportation networks are important complex (but not random)
systems with negligible clustering coefficient. Despite being small in size,
they
are ideal examples for demonstrating the strength of our method for the
analysis of local efficiency. We have made a reinterpretation of the results
obtained for the Boston subway network \cite{latora2001} and performed an
analysis of Barcelona (B) and Madrid (M) underground systems.

The Boston underground transportation system, consisting of $N=124$
stations and $N_l=124$ tunnels, was described both as an unweighted and a
weighted graph in \cite{latora2001}. In the unweighted case, it was found 
that
it
is neither globally nor locally efficient, having $E^{g}=0.1$ and
$E^{l}_{0}=0.006$. This small value for $E^{g}$ gives a false impression of 
low
global efficiency. Although it is only 10 \% of the largest value for fully
connected graph, we should check how much the complex topology of the
Boston subway system improves its efficiency, compared to a regular ring 
with
the same number of sites and links. We found out that such a ring has
$E^{g}_{r}=0.076$, so that the Boston network is by 32 \% more efficient!.
Locally, the original measure relying heavily on the presence of triangles 
of
neighbors has a very small value $E^{l}_{0}=0.006$ \cite{latora2001}, as a
consequence of the typically low clustering in underground transportation
systems. Another comparison could be made against a hub consisting of a
central node of degree $k_c=125$ and 125 peripheral nodes. Such a graph has
the highest possible global efficiency of $E^{g}_{hub} \approx 0.5$ for the
given number of 125 links, but local efficiency ($E^{l}_{0}$ or $E^{l}_{1}$) 
is
zero. Any attempt to locally increase  efficiency of a hub by rearranging 
links,
would eventually lead to a decrease of it on the global scale. Therefore, we
consider that in real systems, such as the Boston subway network, an
appropriate pay-off between global and local efficiencies is achieved. 
Taking
physical distances into account, the global efficiency is increased to
$\tilde{E}^{g}_1=0.63$, while locally remains quite low 
$\tilde{E}^{l}_{0}=0.03$
\cite{latora2001}. Only after the network is extended to include the Boston 
bus
system, it becomes efficient on both scales, with $\tilde{E}^{g}_1=0.72$ and
$\tilde{E}^{l}_{0}=0.46$  \cite{latora2001}. This final result was 
interpreted
as a
small-world behavior. On the basis of our previous discussion of weighted
regular networks it is evident that such an interpretation is not correct. A
simple weighted regular ring with $K=2$ is both globally and locally very
efficient, due to the constant speed of trains and high clustering 
coefficient,
respectively. Weighted efficiencies in real networks with constant speed of
informational transfer are not appropriate measures to give  clear criteria 
for
its
classification. They can only give a hint on up to which extent a particular
real
network can replace the ideal fully connected weighted graph. Furthermore, 
it
seems that only the comparison with the ideal graph is plausible. In most of 
the
cases it is hard to find out what should be the corresponding weighted
"regular" network, because the geographical positions of the nodes in a real
complex network are given and fixed, and usually not equidistant.

In the following analysis of Barcelona and Madrid subways we will include
several technical details and make a step outside a pure theoretical 
research,
offering proposals on how efficiencies of these networks could be improved.
Concerning the Barcelona system \cite{barcelona}, we do not take into 
account
connections by a regular train, but only consider six metro lines. The 
number
of stations and tunnels are $N(\mbox{B})=104$ and $N_l(\mbox{B})=115$,
respectively. When viewed as an unweighted graph, this system has the
average path length of $L(\mbox{B})=9.85$, very small clustering coefficient
$C(\mbox{B})=0.008$ (which is the most important contribution to $E^l_0$),
global efficiency of $E^g(\mbox{B})=0.153$ and a redefined local efficiency 
of
$E^l_1(\mbox{B})=0.080$ (see Table \ref{table2}). Comparing with the
corresponding basic (regular) network with $L_b=23.58$ and $E^g_b=0.095$,
we see that the average path length is more than two times shorter and the
global efficiency improved by 61 \%, due to the complex topology of the
Barcelona system. Furthermore, the local efficiency is nine to ten times 
larger
than if it would have been estimated on the basis of the original equation
\cite{latora2001} or the clustering coefficient. Similar results are 
obtained
for
the Madrid system \cite{madrid}. It consists of 13 metro lines (including 
the
ring
{\it MetroSur}), forming an unweighted network of $N=188$ nodes and
$N_l=223$ links. Due to its larger size, the average path length
$L(\mbox{M})=12.36$ is longer than in the Barcelona system and the global
efficiency is smaller $E^g(\mbox{M})=0.127$, see Table \ref{table2}.
Nevertheless, the values of these two quantities are much better than for 
the
corresponding basic network with $L_b=38.64$ and $E^g_b=0.064$. The
complex topology improves the global efficiency by more than 98 \% ! The
clustering coefficient is larger than in Barcelona ($C(\mbox{M})=0.011$),
because several triangles are formed (particularly around stations {\it Gran
Via}
and {\it Goya}). The higher redefined local efficiency of
$E^l_1(\mbox{M})=0.115$ is a consequence of the larger clustering 
coefficient,
as well as the presence of two rings (metro lines number 6 {\it Circular} 
and
number 12 {\it MetroSur}). Cutting off a reference site belonging to one of
these two rings, its ring-neighbors can still exchange trains along the rest 
of
the ring.

\begin{table}[h]
\caption{Performances of Barcelona and Madrid subway systems.}
\begin{tabular}{|l|rr|} \hline
& Barcelona & Madrid \\ \hline
$N$ & \hspace{1mm} 104 & \hspace{1mm} 188 \\
$N_l$ &  115 &  223 \\
$L$ &  9.85 &  12.36 \\
$C$ &  0.008 &  0.011 \\ \hline
$E^g$ &  0.153 &  0.127 \\
$E^l_0$ &  0.009 &  0.012 \\
$E^l_1$ &  0.080 &  0.115 \\ \hline
$\tilde{E}^{g^{^{^{.}}}}_{1}$ &  0.734 &  - \\
$\tilde{E}^{g}_{2}$ &  0.753 &  - \\
$\tilde{E}^{l}_{0}$ &  0.019 &  - \\
$\tilde{E}^{l}_{1}$ &  0.136 &  - \\
$\tilde{E}^{l}_{2}$ &  0.131 &  - \\ \hline \hline
\end{tabular}
\label{table2}
\end{table}

Similarly to the Boston subway system, the global efficiency of the weighted
Barcelona network is quite high: $\tilde{E}^g_1(\mbox{B})=0.734$ or
$\tilde{E}^g_2(\mbox{B})=0.754$. The main contributions come from several
straight-line subgraphs (such as that between stations {\it Santa Eulalia} 
and
{\it Sagrada Familia}), being identical to fully connected weighted 
subgraphs.
The redefined local efficiency takes values of 
$\tilde{E}^l_1(\mbox{B})=0.136$
or $\tilde{E}^l_2(\mbox{B})=0.131$, that are about seven times larger than
when calculated using the original equation \cite{latora2001}, see Table
\ref{table2}. The efficiencies could be further improved by directly 
connecting
a
few stations that are separated by a long path, although  physically close 
to
each other. Adding only two links, one between stations {\it Can Serra} and 
{\it
Can Vidalet}, and another between {\it Valldaura} and {\it Horta}, two new 
rings
are created. The number of links is increased by only 1.7 \%, while the
increase of the global efficiency is $\delta \tilde{E}^g_1(\mbox{B})= 3.3 
\%$
and that of the local efficiency $\delta \tilde{E}^l_1(\mbox{B})= 26.5 \%$.
Obviously, we can even assume that the stations within these pairs are {\it
connected} or represent a {\it single station}, as we can simply walk from 
one
to the other.

\section{Concluding remarks}
\label{concl}

In this work we focused on two objectives:

1. Introducing a new definition of local efficiency that does not depend
exclusively on the clustering coefficient, and

2. Use that definition, to show that there is another class of complex 
networks
with short average path length and Poisson distribution of degrees, that is 
not
random although its clustering coefficient is negligible.

After accomplishing the first task, we proceeded with a systematic analysis 
of
different types of regular and complex networks. Calculating global and
modified local efficiencies, and taking into account the distribution of
connectivity, we were able to make a clear classification of {\it 
unweighted}
complex networks. The main conclusions that emerge from this study are:

i) The class of small-worlds can be generalized to include systems with
negligible
clustering coefficient. We introduced a new type of networks that has a 
small
number of triangle cycles, but still clearly distinguishable from random 
systems
due to its relatively good local transfer of information.

ii) Small-worlds (both clustered and declustered) as homogeneous systems,
and scale-free networks, as heterogeneous systems, are identified to be both
globally and locally efficient.

Showing that declustered small-worlds behave qualitatively in the same way 
as
standard clustered small-worlds, we addressed today's paradigm of the
importance of the clustering coefficient.

Applying our method to real networks with physical distances and a constant
speed of informational transfer, we found that {\it weighted} efficiencies 
can
be
used only to compare a particular real network with the ideal fully 
connected
weighted graph. As highly clustered weighted regular rings can be both 
globally
and locally efficient, it is hard to establish clear criteria for 
identification
of
small-world behavior. In particular our analysis of the underground
transportation systems of Boston \cite{latora2001}, Barcelona and Madrid
reveals a proper balance between global and local performance. Despite the
constraints on the number of tunnels, global efficiency is noticeably high 
due
to
the complex topology of these networks. On the other hand, allowing for the
use of alternative paths after one station is cut off, the local efficiency
turns to
be five to ten times larger than the results reported in Ref. 
\cite{latora2001}.

\begin{center}
{\bf ACKNOWLEDGMENTS}
\end{center}
Financial support by Fet Open Project COSIN IST-2001-33555 and the
Universities
of Barcelona and Alicante is gratefully acknowledged.


\begin{thebibliography}{99}
\bibitem{lengths}
Explicit formulae for the average path length of  clustered rings with $K=2$ 
and
declustered rings with an
arbitrary $K$ as well as for clusters of the square lattice with periodic
bounday
consitions are reported in Ref. {\protect \cite{VL04}}.
\end{thebibliography}
\end{document}